\documentclass{article}

\usepackage{PRIMEarxiv}

\usepackage[utf8]{inputenc} % allow utf-8 input
\usepackage[T1]{fontenc}    % use 8-bit T1 fonts
\usepackage{url}            % simple URL typesetting
\usepackage{booktabs}       % professional-quality tables
\usepackage{amsmath}        % math
\usepackage{amsfonts}       % blackboard math symbols
\usepackage{nicefrac}       % compact symbols for 1/2, etc.
\usepackage{microtype}      % microtypography
\usepackage{fancyhdr}       % header
\usepackage{graphicx}       % graphics
\usepackage{float}          % [H] for the teaser figure
\usepackage{hyperref}       % hyperlinks; load after float
\graphicspath{{figures/}}

\title{MI-MIDI: Mechanistic Interpretability of Text-to-MIDI Generation Models via Probing, Lenses and Steering}

\author{
  Jakub Poćwiardowski, Mateusz Modrzejewski \\
  Institute of Computer Science \\
  Warsaw University of Technology \\
  Warsaw, Poland \\
  \texttt{jakub.pocwiardowski.stud@pw.edu.pl, mateusz.modrzejewski@pw.edu.pl} \\
}

\begin{document}
\maketitle

\begin{abstract}
Mechanistic interpretability of music generation has concentrated on audio models, leaving symbolic models largely unexplored. We analyze two public text-to-MIDI systems of contrasting design: the purpose-built encoder--decoder text2midi and MIDI-LLM, a Llama~3.2~1B model extended with MIDI tokens using linear probing, the logit and tuned lenses, activation patching and difference-in-means steering. Across these methods, we recover musically meaningful structure and show how architecture shapes its formation and control. Pitch, instrumentation, harmony and texture are linearly decodable in both models. text2midi refines predictions gradually across depth, whereas MIDI-LLM works largely in its inherited textual basis before a sharp late rotation into the musical vocabulary; patching identifies a matching late attenuation of prompt-driven instrument transfer. Steering produces bidirectional changes in register and polyphony in both systems, and in tempo/energy in MIDI-LLM. Our two-orientation protocol isolates directional control and shows that all-layer interventions are robust in text2midi but accumulate disruptively in MIDI-LLM. Together, the results provide a practical toolkit for tracing and controlling musical concepts in symbolic generators. Audio examples are available on a demo website\footnote{\url{https://jpocwiar.github.io/MI-MIDI-Demo/}}.
\end{abstract}


\begin{figure}[H]
    \centering
    \includegraphics[width=0.9\textwidth]{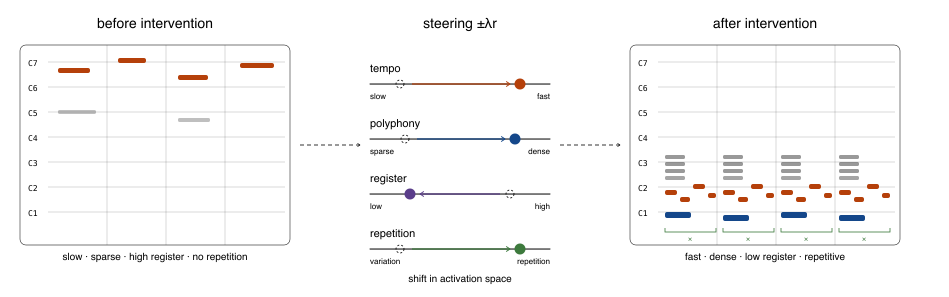}
    \caption{Activation steering as a control panel for text-to-MIDI: shifting residual-stream activations allows steering the generated score. We locate such concepts in two public models and ask when interventions remain safe. }
    \label{fig:teaser}
\end{figure}

\section{Introduction}

Generative models of symbolic music have been present in the literature for a long time, including models that generate scores from text descriptions, yet how they internally represent musical structure remains largely unknown. Mechanistic interpretability has produced a mature toolbox in the language domain with approaches such as probing classifiers~\cite{alain2016understanding,hewitt2019designing}, the logit lens~\cite{nostalgebraist2021logit,belrose2023eliciting}, activation steering~\cite{turner2023steering, subramani2022extracting}, and sparse autoencoders~\cite{huben2024sparse}. Many of these approaches have recently been transferred to text-to-audio music models, especially MusicGen~\cite{vasquez2024exploring,facchiano2025activation,singh2026discovering}. Symbolic models, however, have until very recently been left out of this line of work, even though they offer a methodological advantage: the relations between notes can be easily measured in contrast to audio processing.

This paper fills that gap. We study two public text-to-MIDI models of contrasting architectures: text2midi~\cite{bhandari2025text2midi}, a dedicated encoder-decoder model with REMI+ tokenization, and MIDI-LLM~\cite{wu2025midi}, a general-purpose Llama~3.2~1B language model with a vocabulary extended by MIDI tokens. We ask three questions along the classic interpretability axes: \emph{what} musical information is encoded in their activations, \emph{where and when} the output prediction is formed, and \emph{whether} the encoded information can be used to causally steer generation.

Our main contributions are:
\begin{enumerate}
    \item \textbf{One of the first mechanistic-interpretability studies of symbolic music generation and, to our knowledge, the first centered on text-to-MIDI models}, spanning probing, the logit and tuned lenses, activation patching, and activation steering.
    \item \textbf{A comparative account of prediction formation in two contrasting designs}, revealing gradual refinement in a native encoder--decoder and a sharp textual-to-musical basis transition in a language model repurposed by vocabulary extension.
    \item \textbf{A bidirectional evaluation protocol for identifying robust concept control}, decomposing each steering response into directional and symmetric components.
    \item \textbf{An architecture-dependent recipe for intervention}: targeted single-layer steering remains stable in both models, while text2midi also supports highly specific all-layer steering.
\end{enumerate}

Audio renditions of steered generations are available on the demo website, and code for all experiments will be released.

\section{Related Work}

\paragraph{Text-conditioned music generation.} Music generation is traditionally split into the audio and symbolic domains. Recent attention has focused on text-to-audio models such as Jukebox~\cite{dhariwal2020jukebox}, MusicLM~\cite{agostinelli2023musiclm} and MusicGen~\cite{copet2023simple}. In the symbolic domain, progress was long limited by the lack of captioned data; the MIDICaps dataset~\cite{melechovsky2024midicaps} enabled the first end-to-end text-to-MIDI models, including text2midi~\cite{bhandari2025text2midi} and MIDI-LLM~\cite{wu2025midi}, the two systems studied here.

\paragraph{Mechanistic interpretability.} Interpretability can be understood as the ability to explain a model's decisions in human-comprehensible terms~\cite{doshivelez2017rigorousscienceinterpretablemachine,miller2019explanation}. Of the four senses of \emph{mechanistic} distinguished by Saphra and Wiegreffe~\cite{saphra2024mechanistic}, we adopt the narrowest: the direct study of, and intervention on, a model's internal activations. The specific techniques we build on are linear probing~\cite{alain2016understanding,hewitt2019designing,belinkov2022probing}, the logit lens~\cite{nostalgebraist2021logit} and its trained extension, the tuned lens~\cite{belrose2023eliciting}, and steering via activation additions \cite{subramani2022extracting, turner2023steering}.

\paragraph{Interpretability of music models.} Existing work concentrates on audio models, most of it on MusicGen. V\'asquez et al.~\cite{vasquez2024exploring} decode intermediate layers and steer instrumentation and genre; Facchiano et al.~\cite{facchiano2025activation} steer tempo and brightness with difference-in-means vectors; Zhang et al.~\cite{zhang2024instruct} adapt instruction tuning; Singh et al.~\cite{singh2026discovering} train sparse autoencoders on MusicGen activations; Wei et al.~\cite{wei2024music} probe music-theory concepts on the synthetic SynTheory dataset. For symbolic music, prior interpretability work concerns latent dimensions of VAE-based models~\cite{bryan2023exploring,wang2020learning,pati2019learning}, which does not involve intervention on internal activations of autoregressive generators. Concurrently with this work, Prokopiou et al.~\cite{prokopiou2026latent,prokopiou2026closinglooppidfeedback} steer pitch and duration in an unconditional symbolic transformer with difference-in-means directions. To the best of our knowledge, no prior work applies the broader mechanistic toolbox to text-conditioned symbolic models.

\section{Models and Notation}
\label{sec:models}

\begin{table}[htbp]
    \centering
    \caption{The two studied text-to-MIDI models. Both generate multi-track MIDI from a free-form text description but differ in nearly every design choice.}
    \label{tab:models}
    \begin{tabular}{l l l}
    \toprule
     & \textbf{text2midi}~\cite{bhandari2025text2midi} & \textbf{MIDI-LLM}~\cite{wu2025midi} \\
    \midrule
    Architecture & encoder-decoder & decoder-only (Llama 3.2 1B) \\
    Text conditioning & frozen Flan-T5 encoder + cross-attention & text prompt in context \\
    Analyzed layers & 18 (decoder) & 16 \\
    Hidden size $d_{\mathrm{model}}$ & 768 & 2048 \\
    Tokenization & REMI+~\cite{von2023figaro} & AMT~\cite{thickstun2023anticipatory} \\
    Time encoding & metrical (bars, positions) & absolute (10\,ms grid) \\
    \bottomrule
    \end{tabular}
\end{table}

Two architectural differences organize the analysis: cross-attended encoder memory versus a shared prompt--music residual stream, and metrical REMI+ versus absolute-time AMT tokenization.

Throughout the paper, the activation (residual-stream state) of sample $i$ at layer $\ell$ and token position $t$ is denoted
\begin{equation}
    h_{\ell,t}^{(i)} \in \mathbb{R}^{d_{\mathrm{model}}}.
\end{equation}
Activations are recorded in \emph{predictive alignment}: $h_{\ell,t}$ is the state from which the model predicts token $t$, not the state after consuming it. Causal masking makes a single hooked forward pass over the full sequence equivalent to capturing activations at every generation step.

\section{Linear Probing}
\label{sec:probing}

Generative music models are not explicitly supervised on concepts such as key, chords or intervals, yet it is reasonable to expect that they encode such knowledge internally. We test this with linear probes~\cite{alain2016understanding}: following Hewitt and Liang~\cite{hewitt2019designing}, we deliberately restrict probes to logistic regression, so that high accuracy indicates good linear representation of the concept rather than capacity of the probe itself~\cite{belinkov2022probing}. Unlike prior music probing work on isolated synthetic examples~\cite{wei2024music}, we probe both full, realistic generations, where many concepts interfere, and the controlled SynTheory setting (Section~\ref{sec:syntheory}).

\subsection{Method}

We fit one $L_2$-regularized multinomial logistic-regression probe per layer ($C = 1.0$, L-BFGS, 200 iterations, feature scaling fitted on the training fold only) on activation-label pairs $(h_{\ell,t}, y_t)$, and report accuracy, the majority-class baseline, and their difference, the \emph{lift}. Token-level concepts are probed on a fixed random subsample of 200 sequences, since linear probes saturate well below the full corpus. The sequence-level key variants use all usable sequences. To avoid leakage, all tokens from one sequence are kept in the same cross-validation fold (for SynTheory, all tokens from one sample), so that no probe is tested on a sequence it partly saw in training. We use 5-fold grouped cross-validation stratified by label, and report accuracy as the mean and standard deviation over folds. As a selectivity control~\cite{hewitt2019designing} we retrain each probe on labels shuffled between tokens (group split preserved); the control probe collapses to the majority baseline for every concept (lift $-0.068$ to $+0.001$ for MIDI-LLM, $-0.044$ to $-0.002$ for text2midi), confirming that the accuracies below reflect information in the activations rather than probe capacity.

For sequence-level concepts (estimated key) we also aggregate token activations into one vector per sample, either as the activation of the last labeled melodic token (\textit{last token}) or as the mean over all labeled melodic positions (\textit{mean}).

\subsection{Probing full generations}
\label{sec:probing_full}

\paragraph{Data and labels.} We construct $1\,000$ random music descriptions combining key, meter, tempo, mood and instrumentation, generate one sequence per description by ancestral sampling at temperature $1.0$, with a per-sample seed and a budget of 1024 tokens, and capture activations on all layers. Labeled musical-token positions receive properties derived from the decoded MIDI as listed in Table~\ref{tab:concepts}. Labels are taken from the generated music rather than from the prompt, because generations do not always reliably follow the requested attributes. In particular, the key is estimated from the produced pitch classes with the Krumhansl--Schmuckler method~\cite{krumhansl1990cognitive,cuthbert2010music21}.

\begin{table}[htbp]
    \centering
    \small
    \setlength{\tabcolsep}{5pt}
    \caption{Probed concepts and labels derived from each model's symbolic output. Simultaneity uses a $0.05$\,s tolerance for MIDI-LLM and adjacent REMI positions for text2midi; rhythmic-density windows and bins are tokenizer-specific.}
    \label{tab:concepts}
    \begin{tabular*}{\textwidth}{@{\extracolsep{\fill}} l l @{}}
    \toprule
    \textbf{Concept} & \textbf{Label} \\
    \midrule
    Pitch class            & pitch $\bmod\ 12$ \\
    Octave                 & $\lfloor$pitch$/12\rfloor$ \\
    Instrument family      & General MIDI family of the note's instrument, drums separate \\
    Interval size          & semitone distance to the previous melodic note, binned from unison to octave$+$ \\
    Melodic contour        & direction over the last five pitches: rising, falling, flat \\
    Chord root             & root pitch class of the notes sounding together \\
    Chord quality          & major, minor, diminished, augmented, dominant 7th, other \\
    Harmonic function      & chord root as a scale degree of the estimated key: I, IV, V, other \\
    Texture density        & simultaneous-note count, binned into four classes \\
    Local rhythmic density & local onset count, binned into four classes \\
    Estimated key          & Krumhansl--Schmuckler over the last 20 melodic pitches \\
    \bottomrule
    \end{tabular*}
\end{table}

\begin{figure}[htbp]
    \centering
    \includegraphics[width=0.9\textwidth]{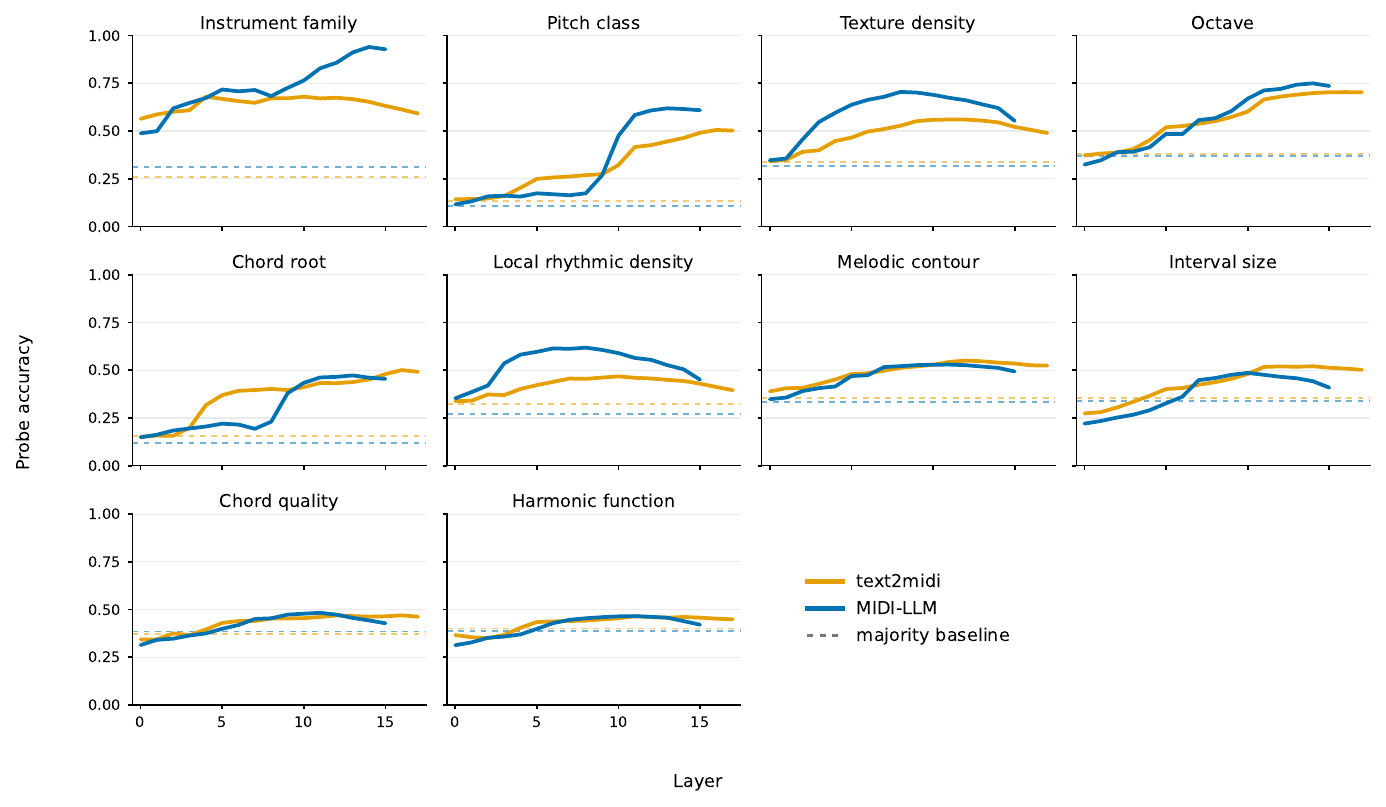}
    \caption{Per-layer probe accuracy for each token-level concept, in text2midi and MIDI-LLM; dashed lines mark each model's majority-class baseline, so a curve's gap above its baseline is the lift. Every best-layer entry in Table~\ref{tab:probing_full} is the peak of the corresponding curve.}
    \label{fig:probing_curves}
\end{figure}

\paragraph{Results.} Table~\ref{tab:probing_full} reports the post-hoc best layer per concept, and Figure~\ref{fig:probing_curves} plots the full per-layer accuracy profile behind each best-layer entry. The results reveal a clear representational hierarchy. Local concepts are strongly decodable in both models: instrument family, pitch class, chord root, octave and texture density reach lifts of $0.36$ to $0.63$ in MIDI-LLM and $0.22$ to $0.42$ in text2midi. Key information is modest at individual tokens (lift $0.06$ to $0.11$) but becomes substantially stronger at last token, when evidence is aggregated across a sequence.

\paragraph{How is key represented?} We probe key at three scales: a single token, the final melodic token, and the mean over all melodic positions. Readout strengthens in that order, showing that key-relevant evidence accumulates across the sequence. A pitch-class histogram control yields lifts of $0.475$ for MIDI-LLM and $0.486$ for text2midi, against $0.545$ and $0.401$ from mean-pooled activations, respectively. It therefore nearly accounts for the MIDI-LLM readout and exceeds the text2midi readout, indicating that much of the sequence-level signal summarizes the note distribution from which the Krumhansl--Schmuckler label is computed. We treat MIDI-LLM's additional $0.070$ lift as suggestive rather than conclusive.

\begin{table}[htbp]
    \centering
    \caption{Probing on full generations: best layer per concept. \textbf{Acc.} is probe accuracy (mean $\pm$ std over 5 CV folds), \textbf{Base} the majority-class accuracy, \textbf{Lift} their difference. Rows sorted by MIDI-LLM lift. Estimated-key labels (bottom rows) are a running Krumhansl--Schmuckler estimate over a sliding window of the last 20 melodic pitches except the \textit{mean} variant (*) which instead pools activations over \emph{all} melodic tokens of the sequence and is labeled by its majority key.}
    \label{tab:probing_full}
    \resizebox{\textwidth}{!}{%
    \begin{tabular}{l cccc cccc}
    \toprule
    & \multicolumn{4}{c}{\textbf{text2midi}} & \multicolumn{4}{c}{\textbf{MIDI-LLM}} \\
    \cmidrule(lr){2-5} \cmidrule(lr){6-9}
    \textbf{Concept} & \textbf{Layer} & \textbf{Acc.} & \textbf{Base} & \textbf{Lift} & \textbf{Layer} & \textbf{Acc.} & \textbf{Base} & \textbf{Lift} \\
    \midrule
    Instrument family      & 10 & $0.680 \pm 0.008$ & 0.260 & 0.419 & 14 & $0.940 \pm 0.009$ & 0.310 & 0.630 \\
    Pitch class            & 16 & $0.505 \pm 0.023$ & 0.134 & 0.372 & 13 & $0.619 \pm 0.019$ & 0.107 & 0.512 \\
    Texture density        & 11 & $0.560 \pm 0.010$ & 0.339 & 0.221 &  8 & $0.705 \pm 0.010$ & 0.319 & 0.386 \\
    Octave                 & 16 & $0.704 \pm 0.010$ & 0.378 & 0.326 & 14 & $0.750 \pm 0.012$ & 0.370 & 0.380 \\
    Chord root             & 16 & $0.501 \pm 0.011$ & 0.155 & 0.346 & 13 & $0.473 \pm 0.024$ & 0.117 & 0.356 \\
    Local rhythmic density & 10 & $0.468 \pm 0.017$ & 0.325 & 0.143 &  8 & $0.618 \pm 0.017$ & 0.271 & 0.348 \\
    Melodic contour        & 12 & $0.551 \pm 0.010$ & 0.354 & 0.197 & 11 & $0.530 \pm 0.012$ & 0.336 & 0.194 \\
    Interval size          & 14 & $0.521 \pm 0.015$ & 0.353 & 0.168 & 10 & $0.486 \pm 0.019$ & 0.339 & 0.147 \\
    Chord quality          & 12 & $0.469 \pm 0.015$ & 0.372 & 0.098 & 11 & $0.483 \pm 0.008$ & 0.385 & 0.098 \\
    Harmonic function      & 11 & $0.465 \pm 0.015$ & 0.401 & 0.065 & 11 & $0.465 \pm 0.009$ & 0.386 & 0.079 \\
    \midrule
    Estimated key (per token)    & 17 & $0.161 \pm 0.012$ & 0.106 & 0.055 & 10 & $0.185 \pm 0.006$ & 0.070 & 0.114 \\
    Estimated key (last token)   & 17 & $0.362 \pm 0.028$ & 0.094 & 0.268 & 11 & $0.336 \pm 0.037$ & 0.073 & 0.263 \\
    Estimated key (mean)*        & 16 & $0.533 \pm 0.012$ & 0.131 & 0.401 & 12 & $0.623 \pm 0.021$ & 0.078 & 0.545 \\
    \bottomrule
    \end{tabular}%
    }
\end{table}

\subsection{Controlled probing on SynTheory in MIDI form}
\label{sec:syntheory}

To complement ecologically valid full generations with isolated musical structure, we run a controlled experiment on SynTheory~\cite{wei2024music}, a synthetic dataset of short examples targeting single music-theory concepts (notes, intervals, scales, chords, chord progressions). Unlike the original audio study, we use the MIDI representation directly, removing the timbre dimension, which leaves $N=1\,080$ samples. For each sample we record the model's activations and train one linear probe per layer with 5-fold cross-validation, with feature scaling fitted on each fold's training split only.

Both models are probed in their native input regime, as required for in-distribution activations: for MIDI-LLM the music tokens are preceded by the official prompt format (an empty, neutral description) and the \texttt{MIDI\_BOS} token, and activations are extracted from musical positions only; for text2midi we use the model's original REMI tokenizer.

\begin{table}[htbp]
    \centering
    \caption{Probing SynTheory (MIDI form, mean pooling, 5-fold CV): best layer per concept. \textbf{Acc.} is the mean accuracy over folds $\pm$ its standard deviation. \textbf{Base} the majority-class accuracy, \textbf{Lift} their difference.}
    \label{tab:syntheory}
    \resizebox{\textwidth}{!}{%
    \begin{tabular}{l cccc cccc}
    \toprule
    & \multicolumn{4}{c}{\textbf{text2midi}} & \multicolumn{4}{c}{\textbf{MIDI-LLM}} \\
    \cmidrule(lr){2-5} \cmidrule(lr){6-9}
    \textbf{Concept} & \textbf{Layer} & \textbf{Acc.} & \textbf{Base} & \textbf{Lift} & \textbf{Layer} & \textbf{Acc.} & \textbf{Base} & \textbf{Lift} \\
    \midrule
    Note pitch class          & 0  & $0.806 \pm 0.072$ & 0.083 & 0.723    & 9 & $0.908 \pm 0.064$ & 0.083 & 0.825 \\
    Register                  & 2  & $0.991 \pm 0.018$ & 0.333 & 0.658    & 1 & $0.991 \pm 0.019$ & 0.333 & 0.657 \\
    Interval                  & 3  & $0.604 \pm 0.035$ & 0.083 & 0.521    & 7 & $0.998 \pm 0.005$ & 0.083 & 0.914 \\
    Scale mode                & 6  & $0.012 \pm 0.015$ & 0.143 & $-0.131$ & 7 & $0.423 \pm 0.051$ & 0.143 & 0.280 \\
    Chord quality             & 10 & $0.375 \pm 0.065$ & 0.250 & 0.125    & 7 & $0.855 \pm 0.067$ & 0.250 & 0.605 \\
    Chord inversion           & 8  & $0.750 \pm 0.050$ & 0.333 & 0.417    & 9 & $0.806 \pm 0.034$ & 0.333 & 0.473 \\
    Chord root pitch class    & 0  & $0.798 \pm 0.052$ & 0.083 & 0.715    & 1 & $0.777 \pm 0.049$ & 0.083 & 0.694 \\
    Chord progression         & 11 & $0.728 \pm 0.056$ & 0.053 & 0.675    & 7 & $1.000 \pm 0.000$ & 0.053 & 0.947 \\
    Progression tonic         & 0  & $1.000 \pm 0.000$ & 0.083 & 0.917    & 0 & $1.000 \pm 0.000$ & 0.083 & 0.917 \\
    \bottomrule
    \end{tabular}%
    }
\end{table}

Each setting is probed in the readout its labels dictate: SynTheory's concepts are properties of a whole example and are read from pooled sequence representations, while on generations the label belongs to the note a token carries and is read per token. They therefore provide complementary views rather than directly comparable lift values. The controlled setting reveals substantial music-theory structure: intervals and chord progressions reach high lift in both models, and chord quality does so in MIDI-LLM.

\begin{figure}[htbp]
    \centering
    \includegraphics[width=0.72\textwidth]{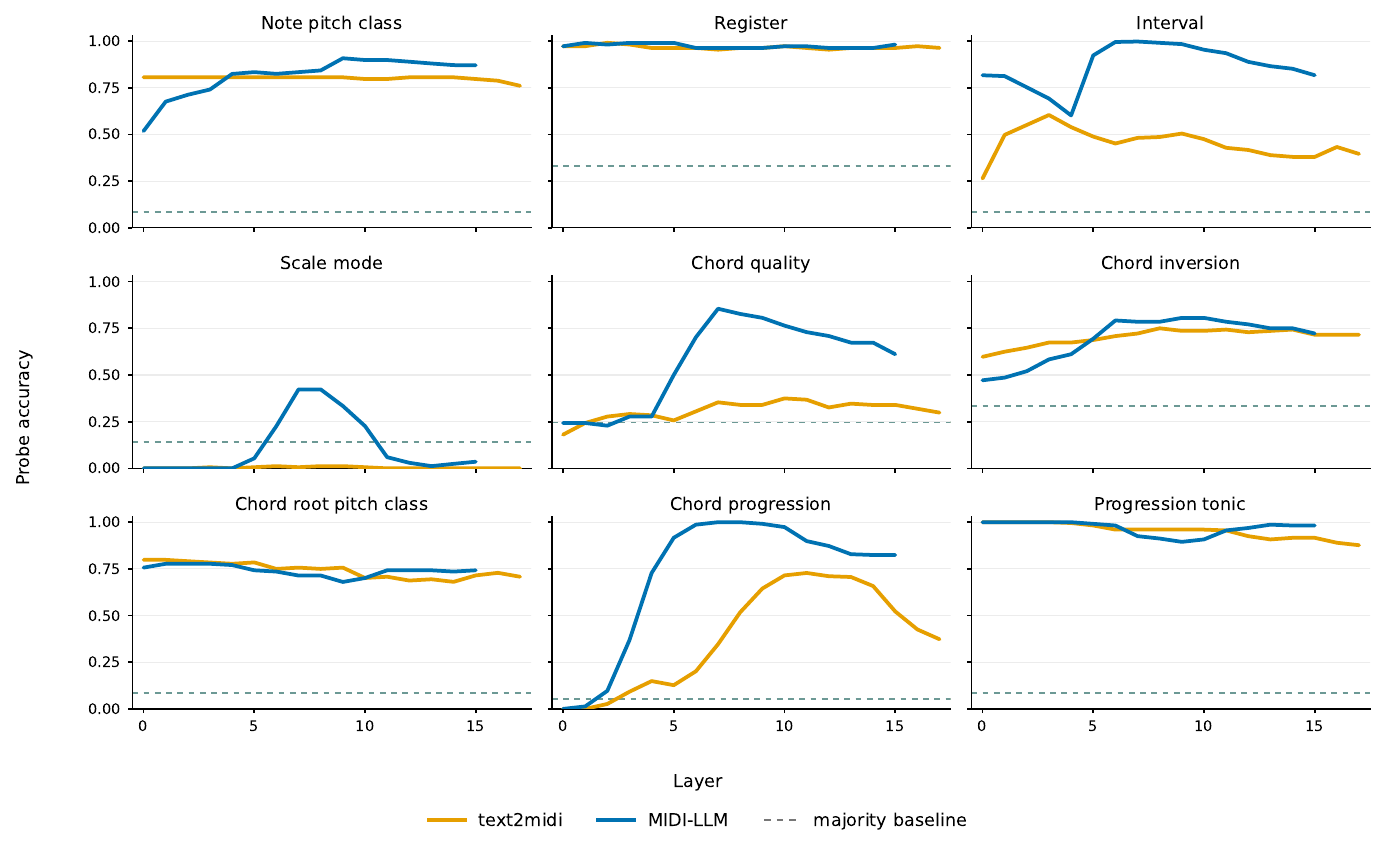}
    \caption{Per-layer probe accuracy on SynTheory (mean pooling) for each concept, in text2midi and MIDI-LLM; dashed lines mark each model's majority-class baseline. Concepts already high at layer~0 (register, progression tonic, pitch-class readouts) are readable from token identity, whereas others (interval, chord quality, chord progression) build up across layers. Every best-layer entry in Table~\ref{tab:syntheory} is the peak of the corresponding curve.}
    \label{fig:syntheory_curves}
\end{figure}

Scale mode provides an additional diagnostic of representational geometry. In SynTheory the seven modes are rotations of one pitch-class set, so the label is fixed by the tonic reference: under the same protocol an absolute-pitch histogram scores at or below chance, while a lowest-note-relative encoding scores perfectly. Against these oracles, text2midi's below-baseline result is consistent with an orderless absolute-pitch representation, whereas MIDI-LLM's positive lift is consistent with a more relative encoding. Figure~\ref{fig:syntheory_curves} shows the full per-layer curves.

\section{Logit Lens and Tuned Lens}
\label{sec:lens}

Probing shows what information is present. The logit lens~\cite{nostalgebraist2021logit} shows when the model's output prediction is formed. Each intermediate activation is passed through the model's final normalization and vocabulary projection, revealing what the model would predict if it stopped at layer $\ell$. We do not assume tied embeddings, instead the architecture's actual output components are used:
\begin{equation}
    z_{\ell,t} = U\left(\operatorname{Norm}\left(h_{\ell,t}\right)\right) + b_U,
    \qquad
    p_{\ell,t} = \operatorname{softmax}\left(z_{\ell,t}\right),
\end{equation}
where $\operatorname{Norm}$ is the final normalization (decoder LayerNorm for text2midi; RMSNorm for MIDI-LLM) and $U$ and $b_U$ are the vocabulary projection ($768 \to$ REMI vocabulary; $2048 \to$ joint text-music vocabulary via \textit{lm\_head}). Since MIDI-LLM's vocabulary mixes text and music tokens, we evaluate the distribution restricted to the musical part. For each layer we report three quantities over the analyzed positions $\mathcal{T}$: the top-1 agreement with the generated token $y_t$, the mean probability the layer assigns to $y_t$, and the entropy of the distribution. Because $y_t$ is the token actually \emph{sampled} during generation rather than the argmax, even the final layer's top-1 agreement stays below $1$. In text2midi the decoder attends to the text encoding at every layer through cross-attention, so we read predictions from successive \emph{decoder} layers. For MIDI-LLM this is the classic decoder-only logit lens. A related decoding of intermediate layers has been applied to MusicGen in the audio domain~\cite{vasquez2024exploring}.

\subsection{Results: gradual refinement vs.\ late binding}

Table~\ref{tab:logit_lens} exposes two qualitatively different regimes. In text2midi, accuracy and correct-class probability grow gradually, with visible jumps around layers 10 to 12, as the decoder repeatedly attends to the encoded description and refines its prediction. MIDI-LLM instead shows a compact transition: its inherited textual basis dominates the classic readout through the first half of the network, followed by a sharp rotation into musical predictions between layers 13 and 14 (vocabulary-mass analysis below). Two cautions apply when reading the columns across models: the depth axes do not start from the same point, since text2midi's first decoder layer already sees the description fully encoded, and the entropies are on different scales, each taken over its own vocabulary of $524$ REMI+ against $55\,026$ musical tokens.

\begin{table}[htbp]
    \centering
    \caption{Logit lens results. \emph{Left:} per-layer agreement with the final token (\textbf{Acc.}), mean correct-class probability (\textbf{Prob.}), entropy, and the MIDI-vocabulary mass share $\operatorname{mass}^{\mathrm{MIDI}}_{\ell}$ (only for MIDI-LLM). \emph{Right:} the layer of peak agreement per predicted token type, with that layer's accuracy and entropy, for text2midi (top) and MIDI-LLM (bottom), sorted by accuracy.}
    \label{tab:logit_lens}
    \begin{minipage}[t]{0.60\textwidth}
        \centering
        \footnotesize
        \setlength{\tabcolsep}{4.5pt}
        \begin{tabular}[t]{c ccc cccc}
        \toprule
        & \multicolumn{3}{c}{\textbf{text2midi}} & \multicolumn{4}{c}{\textbf{MIDI-LLM}} \\
        \cmidrule(lr){2-4} \cmidrule(lr){5-8}
        \textbf{Layer} & \textbf{Acc.} & \textbf{Prob.} & \textbf{Entropy} & \textbf{Acc.} & \textbf{Prob.} & \textbf{Entropy} & \textbf{Mass} \\
        \midrule
        0  & 0.053 & 0.044 & 4.912 & 0.000 & 0.000 & 7.545 & 0.328 \\
        1  & 0.056 & 0.047 & 4.851 & 0.000 & 0.000 & 7.745 & 0.389 \\
        2  & 0.068 & 0.052 & 4.752 & 0.000 & 0.000 & 7.761 & 0.358 \\
        3  & 0.086 & 0.066 & 4.571 & 0.000 & 0.000 & 7.707 & 0.333 \\
        4  & 0.095 & 0.072 & 4.494 & 0.000 & 0.000 & 7.436 & 0.352 \\
        5  & 0.120 & 0.079 & 4.364 & 0.000 & 0.000 & 7.171 & 0.378 \\
        6  & 0.127 & 0.084 & 4.256 & 0.000 & 0.000 & 6.935 & 0.354 \\
        7  & 0.160 & 0.095 & 4.149 & 0.000 & 0.000 & 6.594 & 0.358 \\
        8  & 0.183 & 0.108 & 3.948 & 0.000 & 0.000 & 6.622 & 0.299 \\
        9  & 0.200 & 0.117 & 3.785 & 0.001 & 0.000 & 6.732 & 0.257 \\
        10 & 0.251 & 0.129 & 3.853 & 0.003 & 0.001 & 6.752 & 0.254 \\
        11 & 0.459 & 0.255 & 3.092 & 0.014 & 0.006 & 6.435 & 0.306 \\
        12 & 0.598 & 0.497 & 1.539 & 0.050 & 0.021 & 6.132 & 0.341 \\
        13 & 0.622 & 0.549 & 1.144 & 0.238 & 0.136 & 4.727 & 0.561 \\
        14 & 0.644 & 0.600 & 0.782 & 0.712 & 0.630 & 1.104 & 0.954 \\
        15 & 0.655 & 0.621 & 0.650 & 0.801 & 0.738 & 0.690 & $>$0.999 \\
        16 & 0.666 & 0.631 & 0.643 & --- & --- & --- & --- \\
        17 & 0.677 & 0.608 & 0.915 & --- & --- & --- & --- \\
        \bottomrule
        \end{tabular}
    \end{minipage}\hfill
    \begin{minipage}[t]{0.38\textwidth}
        \centering
        \footnotesize
        \textit{text2midi}\par\smallskip
        \begin{tabular}{l c c c}
        \toprule
        \textbf{Type} & \textbf{Layer} & \textbf{Acc.} & \textbf{Entr.} \\
        \midrule
        PAD       & 0  & 1.000 & 0.170 \\
        TimeSig   & 16 & 0.980 & 0.048 \\
        Duration  & 17 & 0.761 & 0.709 \\
        Program   & 17 & 0.751 & 0.700 \\
        Bar       & 17 & 0.679 & 0.895 \\
        Position  & 16 & 0.634 & 0.652 \\
        PitchDrum & 17 & 0.622 & 0.991 \\
        Velocity  & 17 & 0.604 & 1.171 \\
        Pitch     & 17 & 0.532 & 1.376 \\
        Tempo     & 15 & 0.447 & 0.982 \\
        EOS       & 2  & 0.022 & 5.040 \\
        \bottomrule
        \end{tabular}
        \par\bigskip
        \textit{MIDI-LLM}\par\smallskip
        \begin{tabular}{l c c c}
        \toprule
        \textbf{Type} & \textbf{Layer} & \textbf{Acc.} & \textbf{Entr.} \\
        \midrule
        special  & 15 & 1.000 & 0.040 \\
        time     & 15 & 0.861 & 0.434 \\
        duration & 15 & 0.772 & 0.824 \\
        note     & 15 & 0.768 & 0.816 \\
        \bottomrule
        \end{tabular}
    \end{minipage}
\end{table}

Crucially, the near-zero early-layer values in MIDI-LLM do not imply that early layers are uninformative. On the contrary, the probing results of Section~\ref{sec:probing} show many concepts are linearly decodable there (e.g., local rhythmic density peaks at layer~8). Rather, intermediate representations appear not to be expressed in a basis legible to the final unembedding, a known limitation of the classic logit lens~\cite{belrose2023eliciting}. We test this hypothesis in two ways.

\begin{figure}[htbp]
    \centering
    \includegraphics[width=0.62\textwidth]{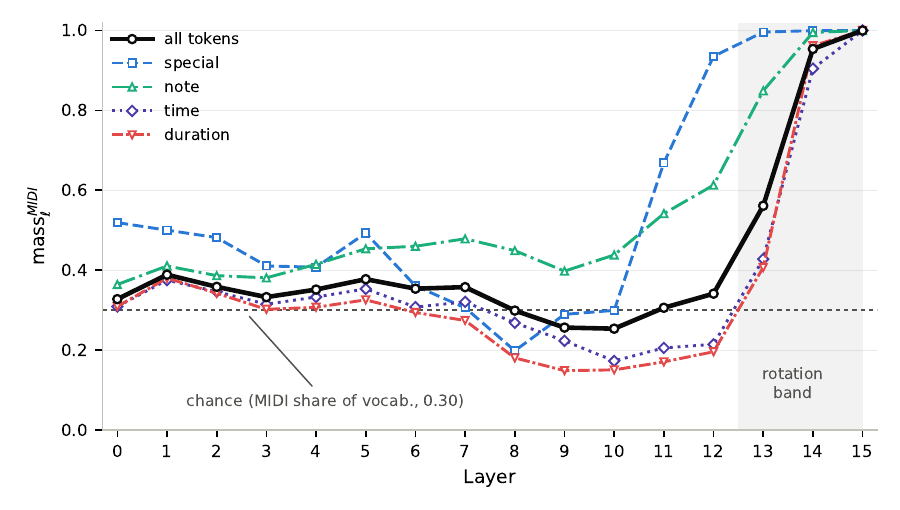}
    \caption{MIDI share of the full-vocabulary probability mass per layer of MIDI-LLM, for all analyzed tokens and by token type. Through layers 0 to 12 the mass stays near the chance share of the MIDI block ($\approx 0.30$, dashed); the readout rotates into the musical vocabulary at layers 13 to 15 (shaded), in the type order special $\to$ note $\to$ time/duration.}
    \label{fig:midi_mass}
\end{figure}

\paragraph{Vocabulary-mass analysis.} If early MIDI-LLM representations remain in an inherited textual basis, then before restricting to $V_{\mathrm{MIDI}}$, the musical vocabulary should receive no more than its size-based share of the full probability mass. We measure that share as
\begin{equation}
    \operatorname{mass}^{\mathrm{MIDI}}_{\ell,t}
    =
    \frac{\sum_{v \in V_{\mathrm{MIDI}}} \exp\!\big(z_{\ell,t}(v)\big)}{\sum_{v \in V} \exp\!\big(z_{\ell,t}(v)\big)}.
\end{equation}
A rotation of $\operatorname{mass}^{\mathrm{MIDI}}$ co-occurring with the accuracy jump at layers 13 to 14 would support this interpretation, and this is what we find (per-layer values in the rightmost column of Table~\ref{tab:logit_lens}, per-type curves in Figure~\ref{fig:midi_mass}). Through the first two-thirds of the network the musical share stays near the size-based share of the MIDI block in the extended vocabulary ($55\,026/183\,282 \approx 0.30$), then rotates almost entirely onto the musical vocabulary across layers 13 to 15, in lockstep with the accuracy jump and entropy collapse. The transition is ordered by token type: special tokens first, note tokens next, and time and duration tokens last, revealing a structured conversion from the inherited basis into an output-ready musical representation.

\paragraph{Tuned lens.} The decisive test is the tuned lens~\cite{belrose2023eliciting}. For each layer $\ell$ a small affine translator $t_\ell : \mathbb{R}^{d} \to \mathbb{R}^{d}$ is trained to map the intermediate state into the final layer's basis, with the frozen normalization and unembedding applied unchanged,
\begin{equation}
    p^{\mathrm{tuned}}_{\ell,t}
    =
    \operatorname{softmax}\Big( U\big(\operatorname{Norm}(t_\ell(h_{\ell,t}))\big) \Big).
\end{equation}
Translators are initialized to identity and trained by distilling the model's own final distribution ($\operatorname{KL}(p_{L,t}\,\|\,p^{\mathrm{tuned}}_{\ell,t})$), on sequences disjoint from the evaluation set ($60\,000$ token positions, four epochs at learning rate $10^{-3}$, with a further $50\,000$ positions held out for evaluation). If the tuned lens lifts early and mid-layer accuracy of MIDI-LLM toward the probing curve, the 13 to 14 jump is an artifact of basis change and information accumulates gradually; if accuracy stays low, the decision genuinely forms late. For text2midi, whose classic lens is already smooth, the tuned lens acts as a control.

\begin{figure}[htbp]
    \centering
    \includegraphics[width=0.72\textwidth]{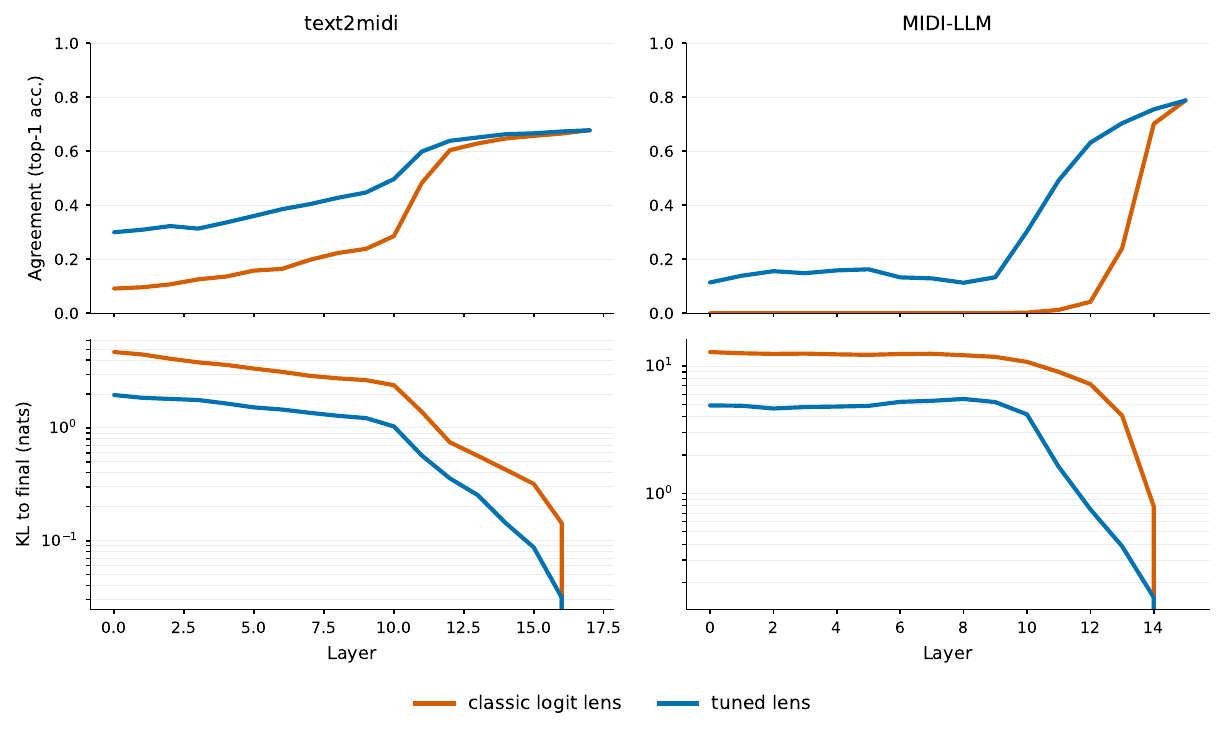}
    \caption{Tuned lens (blue) vs.\ classic logit lens (orange), per layer, for text2midi (left) and MIDI-LLM (right): top-1 agreement with the sampled token (top row) and KL divergence to the final distribution on a log scale (bottom row). For MIDI-LLM the tuned lens makes the prediction readable several layers earlier, smoothing the layer 13 to 14 jump; for text2midi it lifts the whole first two-thirds of the curve without changing its gradual shape.}
    \label{fig:tuned_lens}
\end{figure}

The two explanations contribute in sequence (Figure~\ref{fig:tuned_lens}). In MIDI-LLM, the tuned lens makes the final prediction linearly readable two to three layers earlier than the classic lens suggests and reduces divergence correspondingly sooner, demonstrating that much of the sharp jump reflects a change of basis. Accuracy then continues to rise quickly through the second half, locating genuine prediction refinement after that early readout becomes available. Together with vocabulary mass, this yields a coherent two-stage account: an inherited language-model basis is progressively translated into output-ready musical states while the prediction itself is refined. For text2midi, the tuned lens lifts the early curve without changing its gradual shape, reinforcing the contrast between the architectures.

\section{Activation Patching Across the Binding Transition}
\label{sec:patching}

\begin{figure}[htbp]
    \centering
    \includegraphics[width=0.62\textwidth]{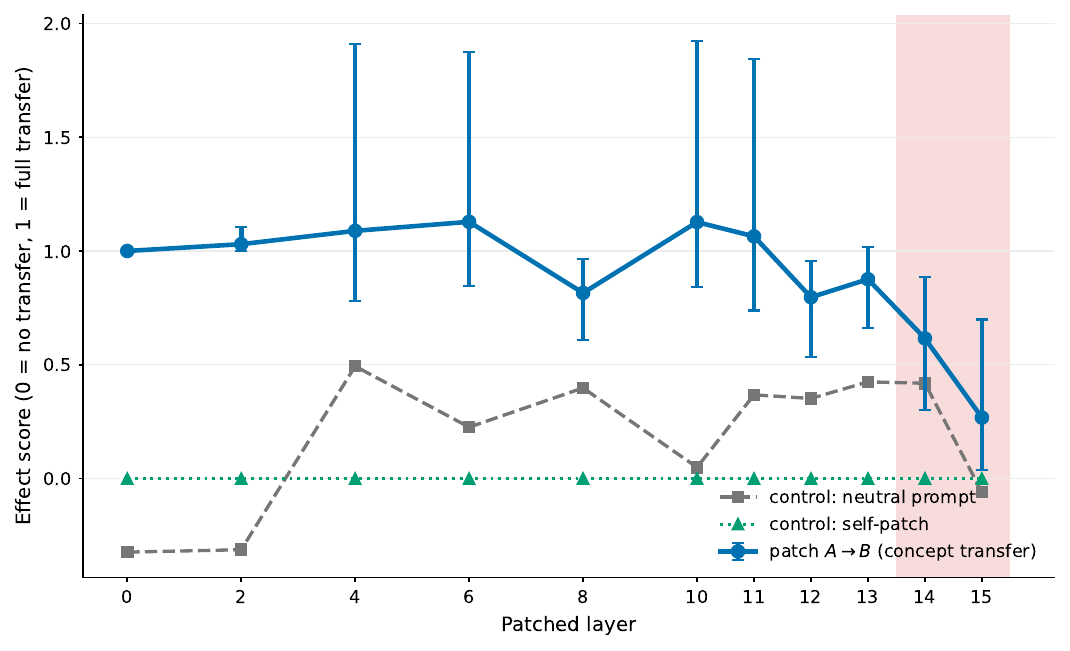}
    \caption{Instrument patching (piano $\to$ violin) at each MIDI-LLM layer input: effect score with paired 95\% bootstrap CIs, the exact-zero self control and the neutral-prompt control. Transfer stays near full through layer 13 and attenuates sharply at layers 14 to 15; the logit-lens transition is shaded.}
    \label{fig:patching}
\end{figure}

Activation patching provides a causal counterpart to the probing and lens results. We generate from one prompt while replacing its prompt activations at the input to layer $L$ with those captured from a contrasting instrument prompt (piano vs.\ violin), and measure how far the piano-note fraction moves from the original toward the contrast ($0$ = no transfer, $1$ = full), guarded by a self-patch control ($0$ by construction) and a neutral-prompt control that catches nonspecific disruption. Each condition uses 10 generations with paired seeds, and confidence intervals are $95\%$ percentiles over $2\,000$ bootstrap resamples of complete seed triplets. In MIDI-LLM, transfer stays near full through layer 13 before falling to $0.62$ and $0.27$ at layers 14 and 15 (Figure~\ref{fig:patching}). This sharp, graded attenuation independently aligns with the late binding transition found by the lenses; residual transfer and shifts in the neutral control make it a transition band rather than a binary boundary.

For text2midi, the mirror experiment patches the cross-attention encoder memory at one decoder layer. The absence of an isolated depth peak relative to the neutral control is consistent with distributed conditioning: the decoder can re-read the original encoder memory at every other layer rather than depending on one bottleneck.

\section{Steering with Activation Additions}
\label{sec:steering}

Control is the strongest test of a mechanistic account. Because the output is symbolic, an intervention's effect for some of the simpler concepts can be measured exactly on the generated score. We therefore report the three concepts that admit such a measure: tempo/energy (note density in tokenizer-native time: notes per beat for REMI+ and per second for AMT), register (mean MIDI pitch), and polyphony (mean number of non-percussive notes sharing an onset), and use the note count as a generation-stability diagnostic. We steer other attributes as well, but they are hard to quantify reliably in multi-track MIDI, so we present those only as audio on the demo page.

\subsection{Difference-in-means directions}
\label{sec:diffmean}

Following Facchiano et al.~\cite{facchiano2025activation}, we compute steering directions as the normalized difference in means between activations of 25 contrastive prompts per pole:
\begin{equation}
    d_{\ell} = \frac{\mu_{\ell}^{B} - \mu_{\ell}^{A}}{\left\lVert \mu_{\ell}^{B} - \mu_{\ell}^{A} \right\rVert_2}.
\end{equation}
The poles are short free-form descriptions designed to contrast the target attribute, with associated musical cues reinforcing each pole, i.e.: \emph{``A very fast and energetic piece, presto tempo, rapid notes''} against \emph{``A very slow and calm piece, largo tempo, long sustained notes''} for tempo/energy, \emph{``A piccolo solo, extremely high and birdlike''} against \emph{``A contrabass solo, extremely low and resonant''} for register, and \emph{``A brass section playing closed-position voicings in rhythmic unison''} against \emph{``A solo trumpet fanfare line with no harmony pads''} for polyphony. Each model is read where its architecture puts the prompt: $\mu_\ell$ is taken at the last prompt token in MIDI-LLM, the only position that has seen the whole prompt~\cite{turner2023steering}, and averaged over a fixed teacher-forced prefix in text2midi, whose prompt reaches every decoder position alike through cross-attention; we call these the models' native readouts.

We add $\alpha d_\ell$ to the residual stream during generation, either at the single layer $L$ where it was computed (\textit{single-layer}) or at every layer (\textit{one-to-all}), and report the resulting change in the concept metric, $\Delta m(\alpha) = m(M_{\alpha}) - m(M_{0})$.

\subsection{Bidirectional evaluation of steering}
\label{sec:sweeps}

A directional intervention has a useful symmetry: its effect should reverse when the steering vector is reversed, whereas nonspecific drift need not. We exploit this by running every configuration in both orientations, $A{-}B$ and $B{-}A$, and jointly regressing the metric on $\alpha$ over run-level points. Fits use the common stable range up to the first $\alpha$ at which the median note count in either orientation drops below half of baseline, and seed-clustered standard errors account for repeated trajectories and pairing across orientations. We call the top of this range $\alpha^{\max}$. Splitting the two fitted slopes, $s_{AB}$ and $s_{BA}$, into their antisymmetric and symmetric parts gives:
\begin{equation}
    s_{\mathrm{dir}} = \frac{s_{BA} - s_{AB}}{2},
    \qquad
    s_{\mathrm{non}} = \frac{s_{BA} + s_{AB}}{2},
    \qquad
    \mathrm{spec} = \frac{|s_{\mathrm{dir}}|}{|s_{\mathrm{dir}}| + |s_{\mathrm{non}}|},
\end{equation}
where $s_{\mathrm{dir}}$ and $s_{\mathrm{non}}$ are the antisymmetric and symmetric components. Specificity $\mathrm{spec} \in [0,1]$ is the antisymmetric share of the response: values near $1$ are consistent with directional control, while values near $0$ indicate symmetric drift.

\subsection{Steering effects}
\label{sec:steering_results}

We apply the protocol to one balanced sweep grid per model: three concepts, both injection strategies, both orientations, nine source layers spanning the depth of each model, $\alpha \in [0,2]$ in steps of $0.25$, and ten seeds per point, making up $9\,720$ sequences per model in the native readout. Table~\ref{tab:steering_effects} shows bidirectional control at the best single-layer configurations: register and polyphony clear the exploratory $2\,\mathrm{SE}$ rule in both models, and tempo/energy does so in MIDI-LLM. Polyphony changes by $20$ to $40\%$, while register moves by one to five semitones across all tracks.

\begin{table}[htbp]
    \centering
    \small
    \setlength{\tabcolsep}{5pt}
    \renewcommand{\arraystretch}{1.08}
    \caption{Steering effects at the best single-layer configuration per concept (Table~\ref{tab:steering_main}): the unsteered metric (\textbf{base}), the metric at the $\alpha$ giving the largest shift within the range where note counts stay intact (\textbf{steered}), and the relative change. ($^*$) response does not clear the exploratory $|s_{\mathrm{dir}}| > 2\,\mathrm{SE}$ rule (text2midi tempo/energy only).}
    \label{tab:steering_effects}
    \begin{tabular}{l l r r r r r r}
    \toprule
    \textbf{Concept} & \textbf{Direction} & \multicolumn{3}{c}{\textbf{text2midi}} & \multicolumn{3}{c}{\textbf{MIDI-LLM}} \\
    \cmidrule(lr){3-5}\cmidrule(lr){6-8}
     & & \textbf{base} & \textbf{steered} & $\Delta$ & \textbf{base} & \textbf{steered} & $\Delta$ \\
    \midrule
    tempo/energy & $\uparrow$ higher & 5.92 & 6.73 ($\alpha{=}2$)   & $+13.6\%^{*}$ & 17.58 & 29.97 ($\alpha{=}1.75$) & $+70.4\%$ \\
    tempo/energy & $\downarrow$ lower & 5.92 & 4.59 ($\alpha{=}1.5$) & $-22.5\%^{*}$ & 17.58 & 17.04 ($\alpha{=}0.75$) & $-3.1\%$ \\
    \midrule
    register & $\uparrow$ higher & 64.49 & 66.34 ($\alpha{=}0.25$) & $+2.9\%$ & 61.14 & 64.33 ($\alpha{=}1.75$) & $+5.2\%$ \\
    register & $\downarrow$ lower & 64.49 & 59.34 ($\alpha{=}1.5$)  & $-8.0\%$ & 61.14 & 58.34 ($\alpha{=}1.5$)  & $-4.6\%$ \\
    \midrule
    polyphony & $\uparrow$ denser  & 3.21 & 4.48 ($\alpha{=}2$) & $+39.6\%$ & 2.84 & 3.46 ($\alpha{=}1$)    & $+21.6\%$ \\
    polyphony & $\downarrow$ sparser & 3.21 & 2.39 ($\alpha{=}2$) & $-25.7\%$ & 2.84 & 2.00 ($\alpha{=}1.75$) & $-29.8\%$ \\
    \bottomrule
    \end{tabular}
\end{table}

Table~\ref{tab:steering_main} decomposes the best configuration per model, concept, and injection strategy. Single-layer interventions remain stable across the full sweep and generally keep symmetric drift small; all single-layer groups except text2midi tempo clear the exploratory $2\,\mathrm{SE}$ rule. text2midi also supports clean \textit{one-to-all} control, while the same strategy in MIDI-LLM reveals an architecture-specific stability limit.

\begin{table}[htbp]
    \centering
    \small
    \caption{Best steering configuration per model, concept, and injection strategy. $|s_{\mathrm{dir}}|$ is the directional strength, $|s_{\mathrm{non}}|$ symmetric drift, spec their antisymmetric share, and $\alpha^{\max}$ the largest stable strength ($\geq 2$: no collapse). The most specific cell clearing the exploratory $|s_{\mathrm{dir}}| > 2\,\mathrm{SE}$ rule is shown (before rounding; layer 0 excluded); (*) means no cell clears it.}
    \label{tab:steering_main}
    \setlength{\tabcolsep}{5pt}
    \renewcommand{\arraystretch}{1.08}
    \begin{tabular*}{\textwidth}{@{\extracolsep{\fill}} l l l c r r c c @{}}
    \toprule
    \textbf{Model} & \textbf{Concept} & \textbf{Strategy} & $L$ & $|s_{\mathrm{dir}}|$ & $|s_{\mathrm{non}}|$ & \textbf{spec} & $\alpha^{\max}$ \\
    \midrule
    text2midi & tempo/energy* & single     & 17 & $0.56 \pm 0.37$  & $0.03$ & 0.94 & $\geq 2$ \\
    text2midi & register      & single     & 9  & $2.17 \pm 0.70$  & $0.87$ & 0.71 & $\geq 2$ \\
    text2midi & polyphony     & single     & 12 & $0.40 \pm 0.16$  & $0.18$ & 0.70 & $\geq 2$ \\
    text2midi & tempo/energy  & one-to-all & 17 & $3.20 \pm 0.50$  & $1.75$ & 0.65 & 1.75 \\
    text2midi & register      & one-to-all & 4  & $3.35 \pm 0.51$  & $0.06$ & 0.98 & $\geq 2$ \\
    text2midi & polyphony     & one-to-all & 16 & $0.55 \pm 0.12$  & $0.03$ & 0.94 & $\geq 2$ \\
    \midrule
    MIDI-LLM & tempo/energy  & single     & 2  & $2.79 \pm 1.28$  & $1.69$  & 0.62 & $\geq 2$ \\
    MIDI-LLM & register      & single     & 8  & $1.52 \pm 0.76$  & $0.06$  & 0.96 & $\geq 2$ \\
    MIDI-LLM & polyphony     & single     & 12 & $0.22 \pm 0.11$  & $0.01$  & 0.96 & $\geq 2$ \\
    MIDI-LLM & tempo/energy  & one-to-all & 7  & $5.13 \pm 1.56$  & $11.90$ & 0.30 & 0.75 \\
    MIDI-LLM & register      & one-to-all & 10 & $6.50 \pm 2.28$  & $18.42$ & 0.26 & 0.75 \\
    MIDI-LLM & polyphony     & one-to-all & 7  & $1.79 \pm 0.77$  & $1.16$  & 0.61 & 0.75 \\
    \bottomrule
    \end{tabular*}
\end{table}

The decomposition turns these responses into an actionable intervention result. In MIDI-LLM, single-layer steering keeps note counts intact and has over fifty times less median symmetric drift than all-layer injection ($0.28$ vs.\ $15.1$; grid medians exclude layer~0), which reaches its stability limit near $\alpha=0.75$. In text2midi, whose decoder re-reads the prompt through cross-attention at every layer, all-layer injection remains stable and yields the most specific directions. Layer~0 provides a further validation of the protocol: large same-way shifts under both orientations are assigned to the symmetric component even when dense note clusters pass the note-count guard. The decomposition is symmetric in $A$ and $B$, so we report magnitudes in Table~\ref{tab:steering_main} and achievable directions in Table~\ref{tab:steering_effects}.

\subsection{Norm-relative steering confirms the architecture effect}
\label{sec:relative}

Because activation norms grow with depth, we test whether MIDI-LLM's all-layer stability limit comes from overscaling its early layers. We make $\alpha$ a share of the local activation size, $h_t \leftarrow h_t + \alpha \lVert h_t \rVert_2\, d$, so that every layer receives an equal relative perturbation, and sweep $\alpha \in [0,0.2]$ in both orientations at one layer per concept (Table~\ref{tab:steering_relative}).

Norm-relative scaling sharpens the same conclusion. Single-layer steering remains intact throughout the full $20\%$ range, whereas all-layer injection reaches its limit at $2\%$ per layer for tempo and $5\%$ for register and polyphony. So the problem is not that a fixed $\alpha$ hits early layers too hard; it is that all sixteen layers are pushed together. For MIDI-LLM, single-layer injection remains the robust choice.

\begin{table}[htbp]
    \centering
    \small
    \caption{Norm-relative steering in MIDI-LLM, decomposed as in Table~\ref{tab:steering_main}. Here $\alpha$ is a fraction of the local activation norm and $|s_{\mathrm{dir}}|$ is per unit of that $\alpha$. ($\geq 0.2$: no collapse.) (*) does not clear the exploratory $|s_{\mathrm{dir}}| > 2\,\mathrm{SE}$ rule.}
    \label{tab:steering_relative}
    \setlength{\tabcolsep}{5pt}
    \renewcommand{\arraystretch}{1.08}
    \begin{tabular*}{\textwidth}{@{\extracolsep{\fill}} l l c r r c c @{}}
    \toprule
    \textbf{Strategy} & \textbf{Concept} & $L$ & $|s_{\mathrm{dir}}|$ & $|s_{\mathrm{non}}|$ & \textbf{spec} & $\alpha^{\max}$ \\
    \midrule
    single     & register       & 8  & $16.1 \pm 7.4$    & $0.03$   & 1.00 & $\geq 0.2$ \\
    single     & polyphony      & 12 & $5.4 \pm 1.6$     & $0.88$   & 0.86 & $\geq 0.2$ \\
    single     & tempo/energy   & 4  & $32.9 \pm 13.2$   & $10.73$  & 0.75 & $\geq 0.2$ \\
    \midrule
    one-to-all & register       & 8  & $140.7 \pm 40.3$  & $46.43$  & 0.75 & 0.05 \\
    one-to-all & polyphony*     & 12 & $37.9 \pm 25.2$   & $22.61$  & 0.63 & 0.05 \\
    one-to-all & tempo/energy*  & 4  & $326.1 \pm 834.3$ & $1\,489.78$ & 0.18 & 0.02 \\
    \bottomrule
    \end{tabular*}
\end{table}

\section{Discussion and Limitations}

\paragraph{Injection strategy.} The results yield an architecture-aware rule for steering. text2midi obtains its most specific directions from all-layer injection, consistent with Facchiano et al.~\cite{facchiano2025activation}, who favor the same strategy on cross-attention-conditioned MusicGen. MIDI-LLM instead responds most robustly to targeted single-layer injection, while repeated additions accumulate in its shared prompt--music residual stream. With one model per architecture we cannot isolate the conditioning pathway from every other design difference, but the contrast provides a concrete hypothesis and a practical strategy to test in future models.

\paragraph{Choosing where to steer.} The most specific source layer is concept-dependent in both models, making empirical layer sweeps valuable; this contrasts with the consistent mid-layer block reported for MusicGen by Facchiano et al.~\cite{facchiano2025activation}. Bidirectional evaluation adds a complementary safeguard to note-count stability by detecting symmetric layer-0 responses that a generation-volume check alone would retain.

\paragraph{What the two probing settings measure.} The two probing settings expose complementary capabilities. SynTheory reveals strong representations of intervals and chord progressions in both models and of chord quality in MIDI-LLM when each concept is isolated. Full generations show which information remains linearly accessible in the richer setting where musical properties co-occur. Their different readouts and distributions answer different questions, so neither is treated as a correction of the other.

\paragraph{Further limitations.} Difference-in-means directions come from text contrasts that also vary lexically and musically, so a direction is only as clean as the contrast behind it. The metrics are proxies for their concepts rather than the concepts themselves: note density stands in for tempo and energy, but rises just as well when more notes sound at once, so the tempo/energy and polyphony numbers are not fully independent. Labels for probing on generations are heuristic estimates from the decoded score, and part of the weakness of the higher-level concepts may be label noise rather than an absent representation. More complex musical concepts, whose unambiguous definition and measurement in multi-track MIDI remain difficult, stay outside the quantitative protocol altogether.

\section{Conclusion and Future Work}

This paper establishes a mechanistic analysis pipeline for text-conditioned symbolic music. Across probing, the logit and tuned lenses, activation patching and steering, we recover internal musical structure, trace two distinct routes by which predictions are formed, causally transfer prompt conditioning, and produce bidirectional changes in register and polyphony in both models and tempo/energy in MIDI-LLM. The architecture comparison also yields a practical intervention rule: all-layer steering suits cross-attention-conditioned text2midi, while targeted layers provide robust control in MIDI-LLM. Our bidirectional protocol makes these effects measurable by separating directional response from symmetric drift. Moving beyond hand-chosen concepts, ongoing work trains sparse autoencoders~\cite{huben2024sparse,elhage2022toy} on both models to discover and characterize musical features unsupervised. Those results will appear in a follow-up publication.

\section*{Acknowledgments}
We gratefully acknowledge Polish high-performance computing infrastructure PLGrid (HPC Centers: ACK Cyfronet AGH) for providing computer facilities and support within computational grant no. PLG/2025/018892.

%Bibliography
\bibliographystyle{unsrt}
\bibliography{references}

\end{document}